\documentclass{iaus}
\usepackage{graphicx}
\usepackage{epsf}
\newcommand{\masyr}{ \ {\rm{mas \ yr^{-1}}}\>}
\newcommand{\kms}{ \ {\rm{km \ s^{-1}}}\>}
\newcommand{\PM}{{\rm PM}}

\title[Proper Motions of the Magellanic Clouds] 
{New Analysis of the Proper Motions of the Magellanic Clouds 
using \textit{HST}/WFPC2}

\author[Kallivayalil \etal ]   
{Nitya Kallivayalil$^1$, Roeland P.van der Marel$^2$, Jay
  Anderson$^2$, Gurtina Besla$^3$ \and Charles Alcock$^3$}
\affiliation{$^1$Pappalardo Fellow, MIT Kavli Inst. for
  Astrophysics \& Space  Research, 70 Vassar Street, Cambridge,
  MA, 02139, USA; email: {\tt nitya@mit.edu} 
\\[\affilskip]
$^2$Space Telescope Science Institute,  3700 San Martin Drive, Baltimore, 
MD 21218, USA
\\[\affilskip]
$^3$Harvard-Smithsonian Center for Astrophysics,  60 Garden Street,
Cambridge, MA 02138, USA}

\pubyear{2008}
\volume{256}  
\pagerange{119--126}
\jname{The Magellanic System: Stars, Gas, and Galaxies}
\editors{Jacco Th. van Loon \& Joana M. Oliveira, eds.}
\begin{document}

\maketitle

\begin{abstract}
In HST Cycles 11 and 13 we obtained two epochs of ACS/HRC data for fields
in the Magellanic Clouds centered on background quasars. We used these
data to determine the proper motions of the LMC and SMC to better than
5\% and 15\% respectively.  The results had a number of unexpected
implications for the Milky Way-LMC-SMC system. The implied
three-dimensional velocities were larger than previously believed and
close to the escape velocity in a standard $10^{12}$ solar mass Milky Way
dark halo, implying that the Clouds may be on their first
passage. Also, the relative velocity between the LMC and SMC was
larger than expected, leaving open the possibility that the Clouds may
not be bound to each other. To further verify and refine our results
we requested an additional epoch of data in Cycle 16 which is being
executed with WFPC2/PC due to the failure of ACS. We present the
results of an ongoing analysis of these WFPC2 data which indicate good
consistency with the two-epoch results.

\end{abstract}

\firstsection 
\section{Introduction}
The Large and Small Magellanic Clouds (LMC \& SMC) at distances of
$\sim 50$ kpc from the Sun, and $\sim 25$ kpc from the Galactic Plane,
provide one of our best probes of the composition and properties of
the Galactic dark halo, and have long been upheld as the poster-child
for a strongly interacting system, both with each other and with the
Milky Way (MW). It has commonly been assumed that the Clouds have made
multiple pericentric passages about the MW, and indeed current
formation theories for the Magellanic Stream, which may involve tidal
or ram-pressure forces, require multiple pericentric passages in
order to be viable stripping mechanisms (\cite[Gardiner \& Noguchi
1996]{GN96}; hereafter GN96, \cite[Connors \etal \ 2006]{Connors06},
\cite[Mastropietro \etal \ 2005]{Mastropietro05}, \cite[Yoshizawa \&
Noguchi 2003]{YN03}, \cite[Lin \etal \ 1995]{Lin95}, \cite[Moore \&
Davis 1994]{MD94}, \cite[Heller \& Rohlfs 1994]{HR94}, \cite[Lin \&
Lynden-Bell 1982]{LL82}, \cite[Murai \& Fujimoto 1980]{MF80}).

However, recent high-precision proper motion measurements for the
Clouds made by our group with two epochs of ACS High Resolution Camera
(HRC) data in Cycles 11 and 13, where we measured the proper motion of
LMC stars relative to background quasars, imply that the LMC
tangential velocity is $\sim 370 \kms$, approximately $100 \kms$
higher than previously thought (\cite[Kallivayalil \etal \ 2006a]{K1},
\cite[Kallivayalil \etal \ 2006b]{K2}; hereafter K1 \& K2). The proper
motion values of GN96, which have been adopted in all theoretical
models of the formation of the Stream thus far, are not consistent
with the new HST result. In particular, for the LMC there is a
7-$\sigma$ difference. The values for the SMC are in more acceptable
agreement (3-$\sigma$ difference). The new measurements also indicate
a significant relative velocity between the LMC \& SMC of $105 \pm 42$
${\rm km \ s^{-1}}$. This has been assumed to be closer to $\sim 60
\kms$ in theoretical models, i.e., approximately the value for the SMC
to be on a circular orbit around the LMC.

These results have surprising physical implications which require a
reconsideration of the formation mechanism for the Stream. These
include the possibility that the LMC may only be on its first passage
about the MW (\cite[Besla \etal \ 2007]{Besla07}; hereafter B07). B07
demonstrated this by studying the past orbital paths of the LMC using
our observed proper motions and errors in a $\Lambda$CDM-motivated
dark halo with a NFW profile (\cite[Navarro \etal \
1996]{Navarro96}). This gave rise to starkly different trajectories
for the LMC than those produced in a simple isothermal halo potential:
even in the `best case' scenario (proper motion in the west direction,
$\mu_W, \ +4 \sigma$), the LMC only completes 1 orbit within 10 Gyr
and reaches an apogalacticon distance of 550 kpc (see Figure~4 in
B07). Subsequently, this has led to a series of papers exploring
whether the LMC is indeed \textit{bound} to the MW (e.g., \cite[Shattow \&
Loeb 2008]{SL08}, \cite[Wu \etal \ 2008]{Wu08}). Perhaps even more
provocative is the possibility that that the LMC \& SMC may have only
recently become a binary system (K2, although see Besla \etal \ these
proceedings).

Such large motions were unexpected in light of our previous
understanding of the MW-LMC-SMC system, and it is therefore crucial
that they be verified and further improved through the acquisition of
additional data.  Because of the large distances involved, even small
differences in the proper motions can give vastly different orbits for
the Clouds ($1 \masyr$ $\approx 238$ ${\rm km \ s^{-1}}$ at the
distance of the LMC). We thus applied for and were successful in
getting a third epoch of snapshot imaging for our quasar-fields in
Cycle 16. These executed with WFPC2 due to the failure of ACS. In this
paper we present the preliminary results of an on-going analysis of
these WFPC2 data. In \S~2 we present the WFPC2 data and analysis
strategy with special attention to the relative size of the position
errors vis-a-vis ACS. We describe the main sources of systematic error
in both the ACS and WFPC2 data. In \S~3 we present results based on
simple cuts aimed at minimizing these systematic errors, and discuss
the expected overall improvement that this additional epoch
affords. Since our analysis is still in the process of being refined,
we present the results only in comparative fashion, both to our
ACS-only two-epoch results (K1 \& K2), and to those of \cite[Piatek
\etal \ 2008]{Piatek08} (hereafter P08), who recently re-analyzed our
ACS data using their own methods to obtain results that are consistent
with ours. A brief summary and future prospects are presented in \S~4.

\section{WFPC2 Data \& Analysis}
\subsection{Description of Observations}
In our first epoch program (Cycle 11, PI: Alcock), we imaged fields
around 40 quasars behind the Magellanic Clouds (\cite[Geha \etal \
2003]{Geha03}) with the ACS High Resolution Channel (HRC) in snapshot
mode (54 targets had been approved and we achieved a 74\% completion
rate). Each field was imaged in the $V$-band (F606W) with an
eight-point dither pattern, to perform astrometry. We also obtained
two $I$-band (F814W) exposures for each field, to allow the
construction of color-magnitude diagrams (CMDs) and investigate any
color-dependent effects.  For the second epoch (Cycle 13, PI: Alcock)
we proposed the same observational strategy (except without
F814W). For the third epoch (Cycle 16, PI: Kallivayalil), however, due
to the recent failure of ACS we proposed to use the WFPC2/PC.  We
requested the 40 targets observed in the first epoch in snapshot mode
again. Our observational approach is similar to that in epochs 1 \&
2. We used the $V$-band (F606W) filter on the Planetary Camera (PC)
and a 5 or 6-point dither pattern.  The first concern in switching
from the HRC to the PC was one of sensitivity. However our target
quasars are relatively bright, ranging from $16.4 \le V \le 20$. The
astrometric error is inversely proportional to the $S/N$ for the
quasar. Aiming at $S/N\sim100-200$ with the F606W filter yielded total
science exposure times ranging from 2.8 to 20 minutes, thereby making
these attractive snapshot targets.  New data are still being obtained:
so far we have obtained 16 out of an expected $\sim 20$ targets, 13 in
the LMC and 3 in the SMC.

In brief, our goal is to measure the motion of the quasar relative to
the L/SMC stars, which should all be moving together. The L/SMC stars
will define a reference frame against which we measure the motion of
the quasar; the L/SMC's motion is then the reflex of this motion.

\subsection{Astrometric Precision: Random Errors}
\begin{figure*}
\centerline{%
\epsfxsize=0.4\hsize
\epsfbox{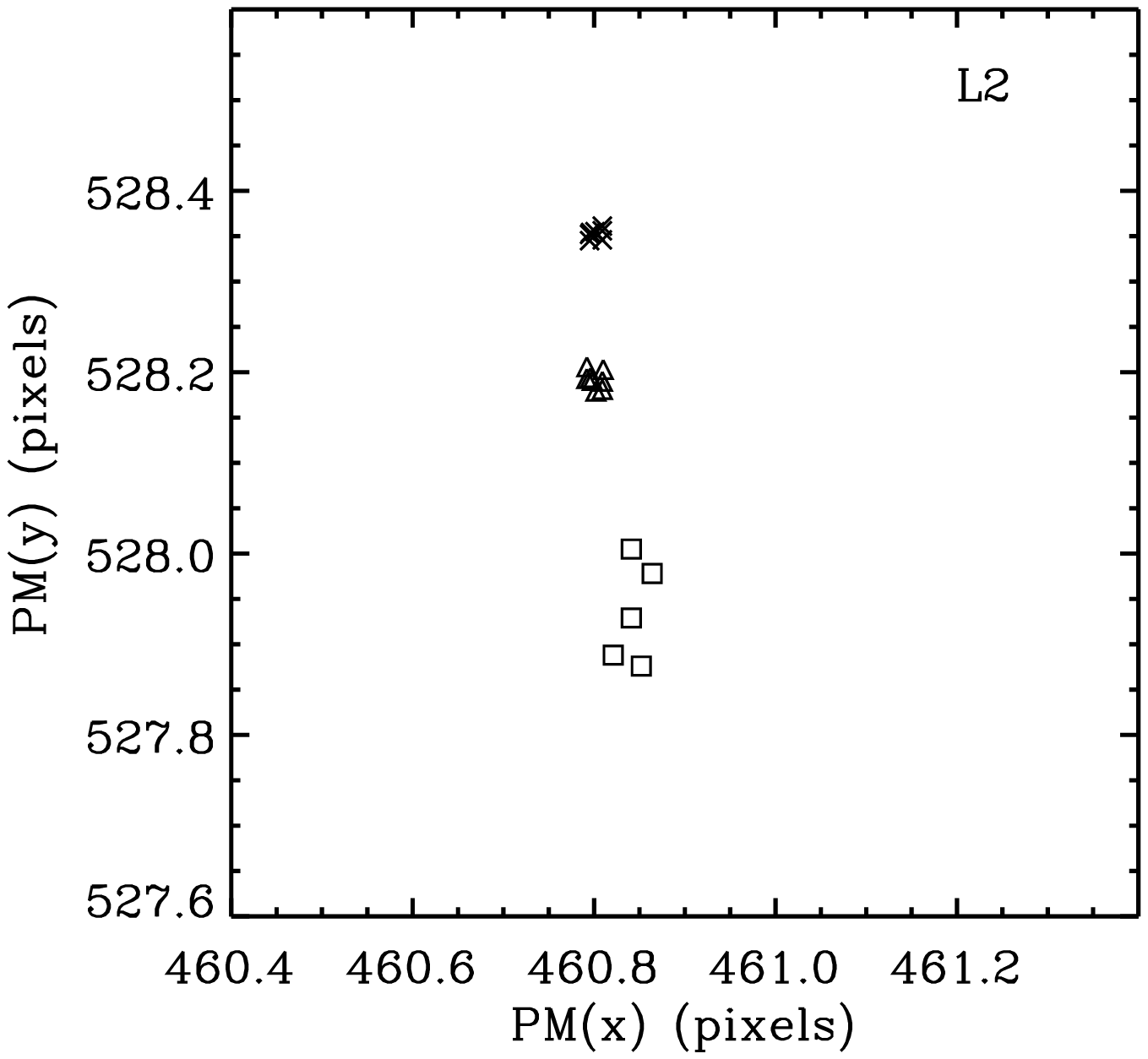}
\epsfxsize=0.4\hsize
\epsfbox{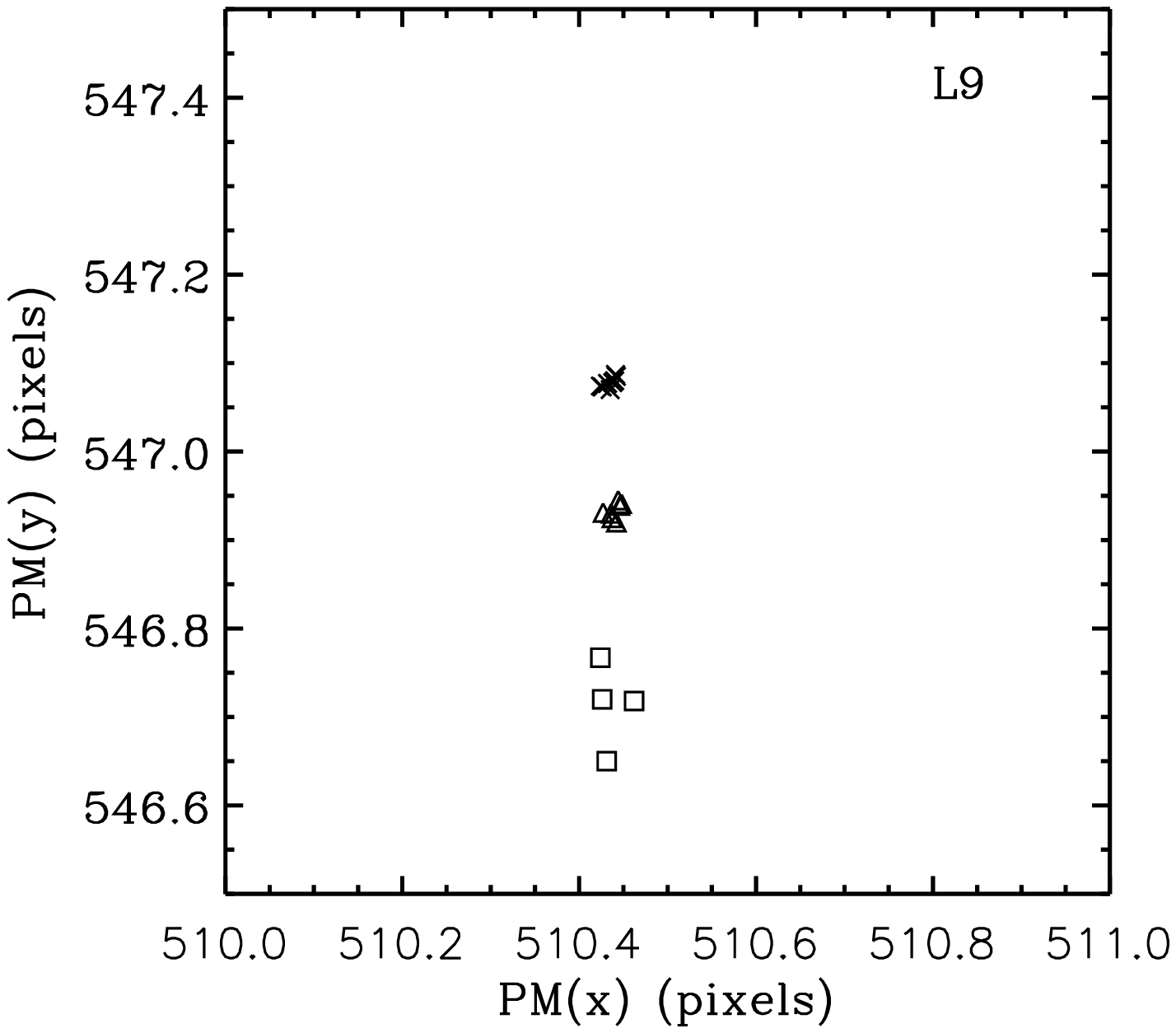}}
\vskip -0.7truecm
\centerline{%
\epsfxsize=0.4\hsize
\epsfbox{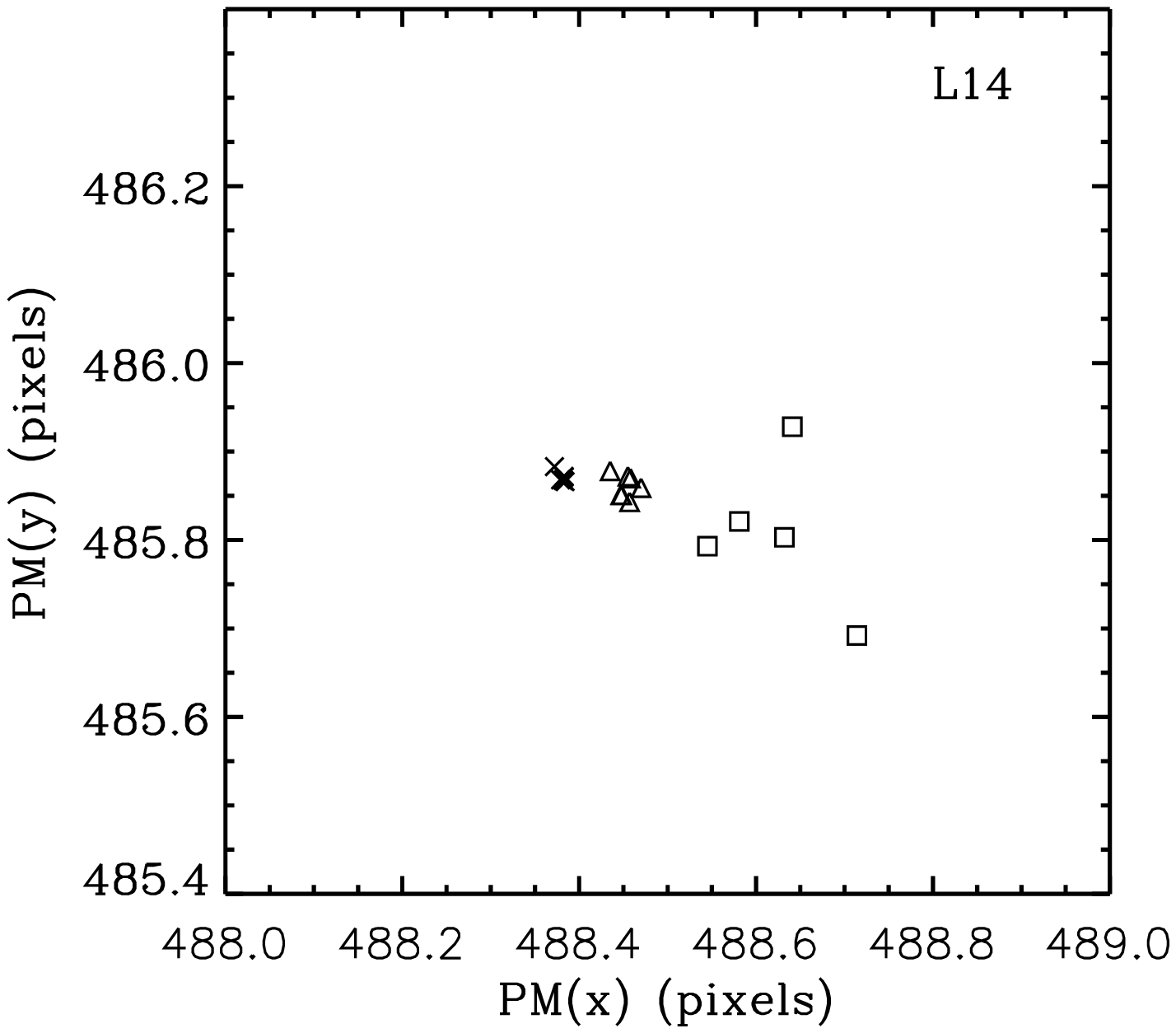}
\epsfxsize=0.4\hsize
\epsfbox{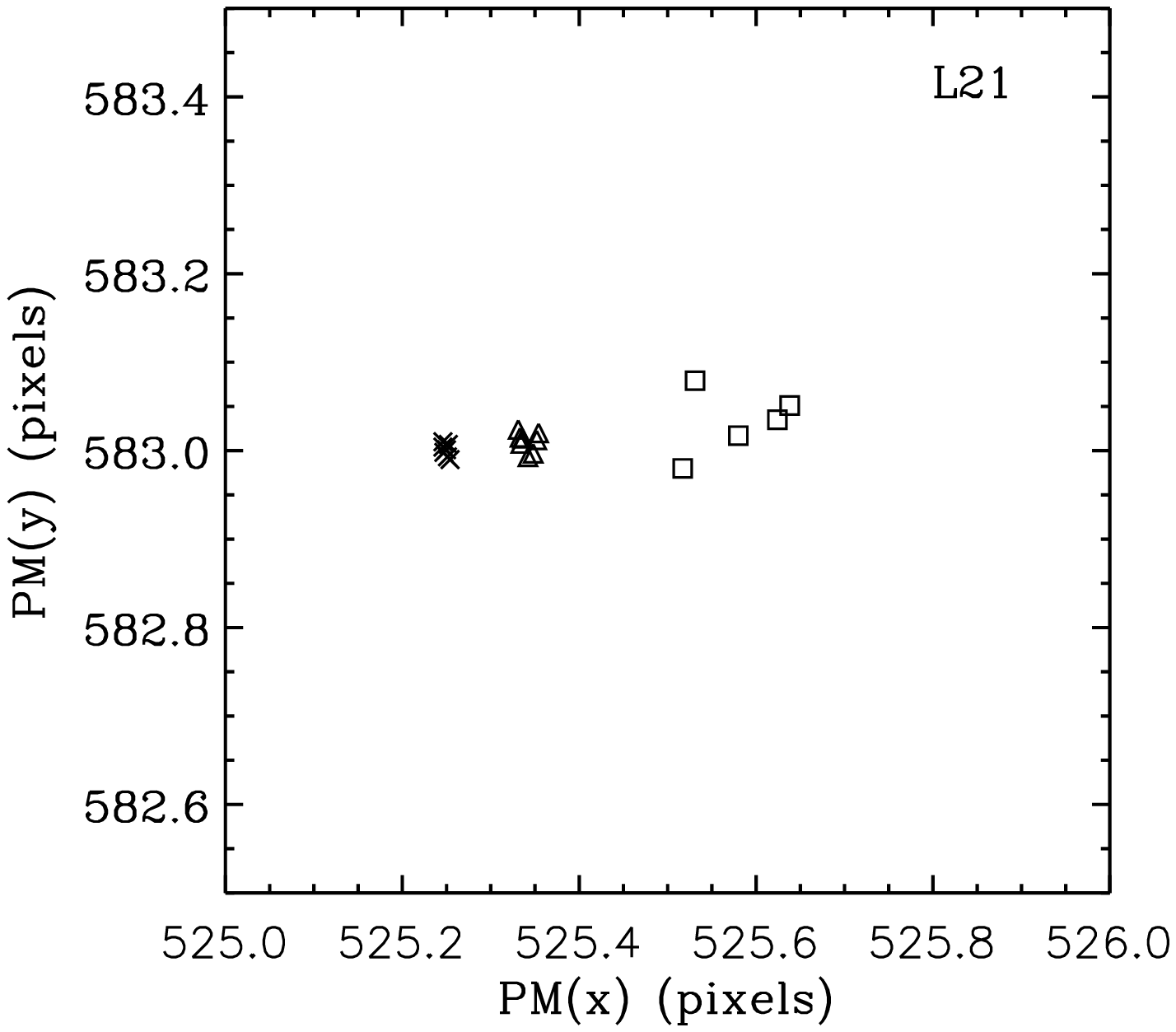}}
\caption{Positions, in a reference frame based on the 1st epoch
  images, of the quasar over time for 4 randomly chosen LMC fields
  showing ACS epoch 1 (\textit{crosses}), ACS epoch 2
  (\textit{triangles}) and PC epoch 3 (\textit{open squares}). The scatter
  per epoch gives the relative size of the RMS errors in the position
  of the quasar relative to the star-field, given that the observation
  has been repeated many times per epoch.}
\label{random_errors}
\end{figure*}

There are many stars in each field, so the random errors in our
transformations will be very small. The astrometric accuracy is
therefore dominated by the accuracy with which we can measure the
position of the quasar in each field relative to the surrounding
stars. In Figure~\ref{random_errors} we show the position of the
quasar over time, in a reference frame based on the 1st epoch images,
for 4 randomly chosen LMC fields (see K1 for definitions of these
fields), showing ACS epoch 1 (\textit{crosses}), ACS epoch 2
(\textit{triangles}) and PC epoch 3 (\textit{open squares}). The RMS
error in the position of the quasar is roughly 3 times as large for
WFPC2 as for ACS ($\sim 0.021$ HRC pixels versus 0.008 HRC pixels).

\subsection{Astrometric Precision: Sytematic Errors}

\begin{figure*}
\begin{center}
\epsfxsize=0.7\hsize
\epsfbox{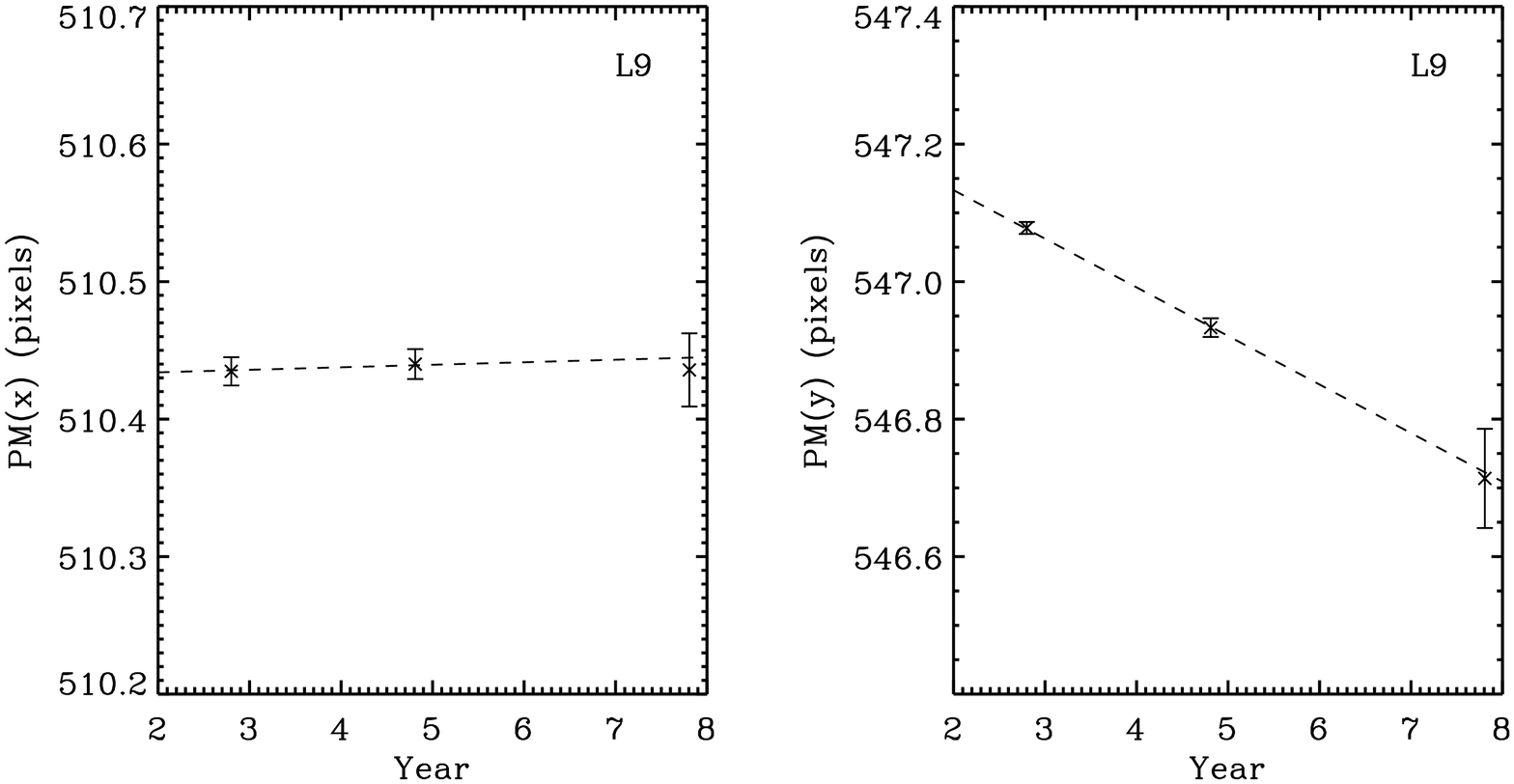}
 \caption{Geometrically-corrected positions ($x$ on the left, $y$ on the
  right) for the QSO in field L9 for the three epochs of data. The
  line connects the first two epoch ACS measurements. The third epoch
  WFPC2 measurement is consistent with this.}
   \label{sys_errors}
\end{center}
\end{figure*}

One of the main drivers in applying for a third epoch was to verify
that there were no residual systematic effects in our measurements in
addition to those that we already accounted for in our error-bars
(e.g., residual geometric distortion solution errors). The third epoch
does give us a handle on this as follows: there is always a straight
line through two measurements, but if the third measurement doesn't
line up then that is a clear indication of some systematic effect.
Figure~\ref{sys_errors} shows geometrically-corrected coordinates for
the quasar as a function of time, in the frame of the 1st-epoch, for
field L9 for all three epochs. This is essentially a measure of the
reflex proper motion of the field. The line connects the two ACS
epochs and is not a fit. Thus even without detailed calculations it is
easy to see that the third epoch is lining-up well, which rules out
the presence of any \textit{major} residual systematic errors.

We do know that there are astrometric errors due to
charge-transfer-efficiency degradation (CTE) as well as other
magnitude-dependent effects, mostly along the detector $y$-axis for
ACS and along both $x$ and $y$ for the PC. These trends are small : $<
0.02$ pixels for ACS, and $< 0.03$ pixels for PC but are significant
when considering the accuracy we are aiming for. Our strategy of
observing our $N$ fields at random roll-angles of the telescope means
that detector-frame-based errors average to zero as
$N^{-1/2}$. However, they do introduce scatter. P08 showed
that by accounting for these effects the RMS scatter between fields
can be reduced. Our current analysis also indicates that explicitly
calibrating these effects can improve random errors.


\section{Results}
\begin{figure*}
\begin{center}
 \includegraphics[width=3.5in]{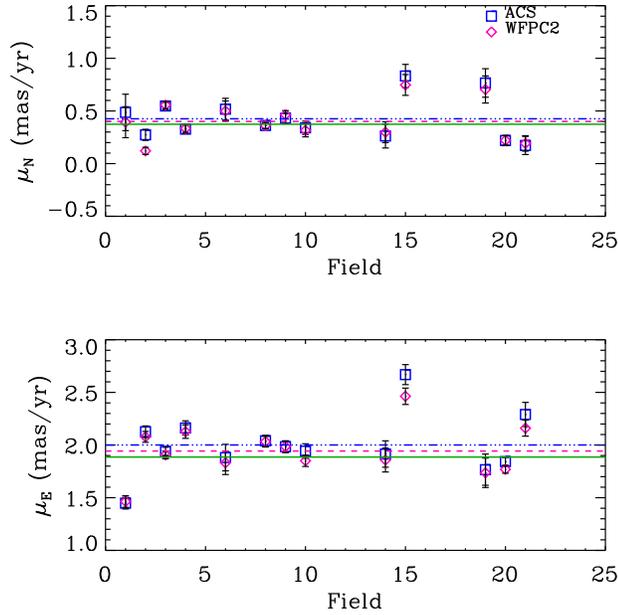}
\caption{Field-by-field comparison for $\mu_N$ (\textit{top}) and
  $\mu_E$ (\textit{bottom}) for the LMC (see K1 for a definition of
  field numbers). Blue squares indicate two-epoch values, pink
  diamonds show three-epoch values. Lines are straight averages,
  dot-dashed for two-epoch, dashed for three-epoch and solid for the
  P08 analysis of these fields from our two-epoch data.}
\end{center}
\label{comp}
\end{figure*}

Since our analysis is still being fine-tuned, we focus in this section
on demonstrating that there is generally good consistency between
results from the third epoch of data and those from the first two
epochs, once some simple cuts in magnitude-space to minimize
systematic effects have been adopted (analogous to those in
P08). Figure~\ref{comp} is a field-by-field comparison for the LMC of
the two-epoch proper motion results (\textit{blue squares}) and the
three-epoch results (\textit{pink diamonds}). Proper motion in the
north direction, $\mu_N$, is shown on top and proper motion in the
east direction, $\mu_E$, is shown on the bottom. Lines show simple
averages for the two-epoch case (\textit{dot-dashed}), the three-epoch
case (\textit{dashed}), and for the P08 analysis of the first two
epochs of these fields (\textit{solid}). There is generally good
agreement to within $1-\sigma$.


\begin{figure*}
\vskip-3.0truecm
\centerline{%
\includegraphics[width=.6\textwidth]{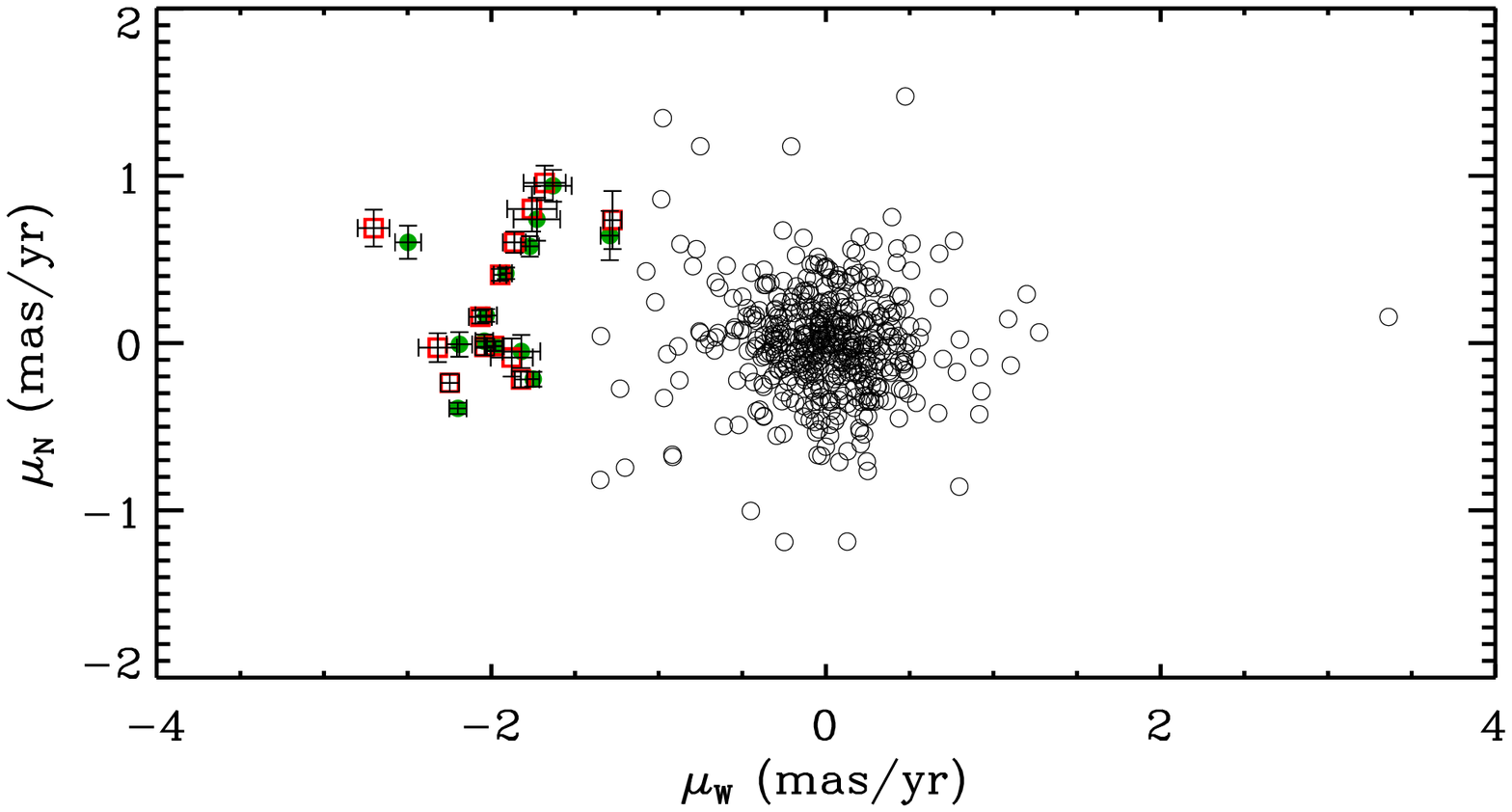}
\includegraphics[width=.6\textwidth]{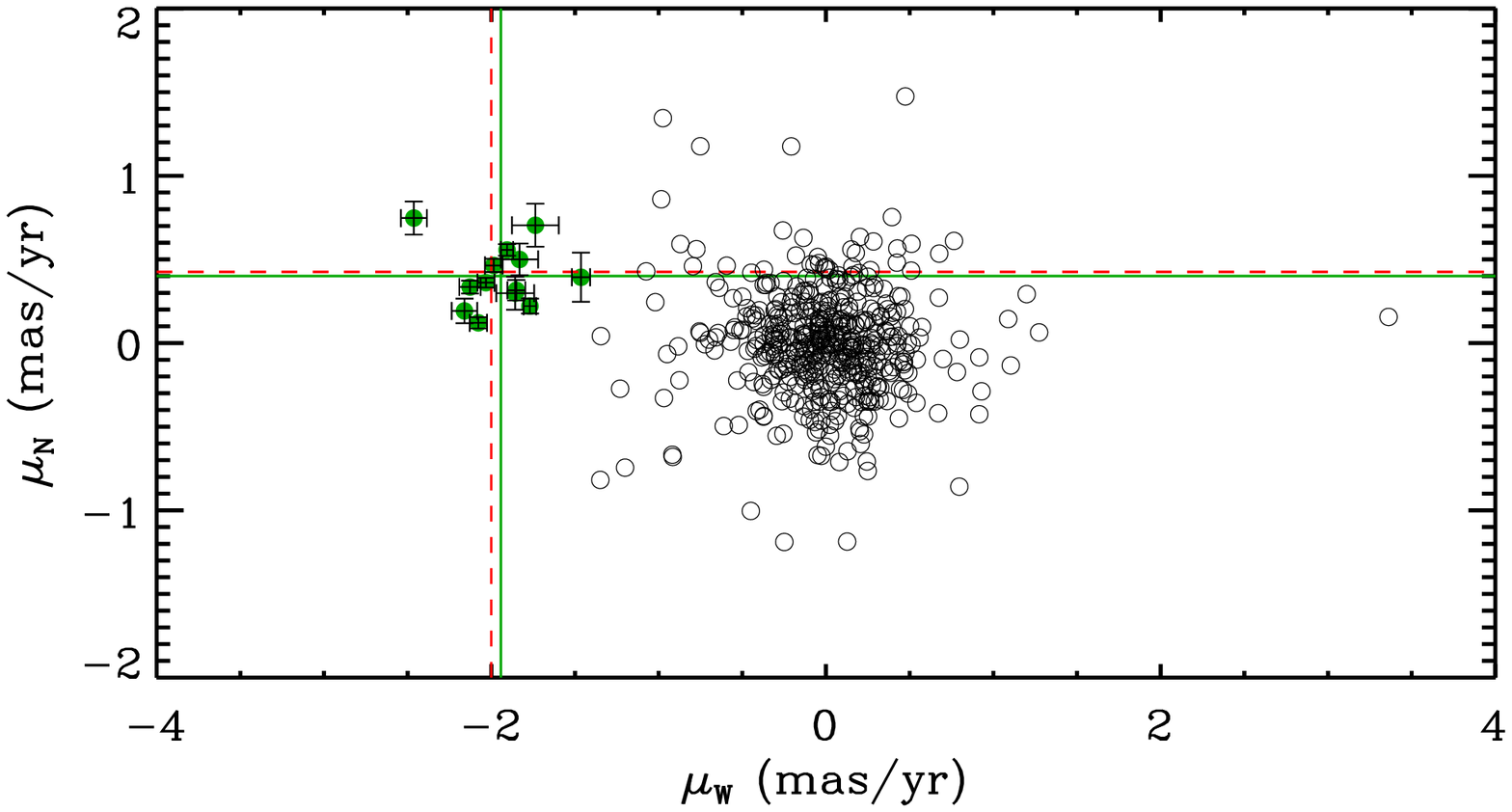}}
\vskip-1.4truecm
\caption{(\textit{Left}) The observed PM ($\mu_{W}$, $\mu_N$) for all
  the quasar fields with three epochs of data (\textit{filled
  circles}). Over-plotted are the corresponding values from the
  two-epoch analysis (\textit{squares}). (\textit{Right}) The estimates
  $\PM_{\rm est}({\rm CM})$ of the LMC center of mass proper motion
  after corrections for viewing perspective and internal rotation have
  been made for the three-epoch fields. The residual PMs of the LMC
  stars in all the fields are plotted with open circles in both
  panels. The reflex motions of the quasarss clearly separate from the
  star motions.  The straight lines mark the averages, dashed for the
  two-epoch analysis and solid for the three-epoch one.}
\label{PMs}
\end{figure*}

The proper motion values for the LMC quasar-fields with three epochs
of data are shown in the ($\mu_W,\mu_N$)-plane in
Figure~\ref{PMs}. They are compared to the residual proper motions of
all the LMC stars that were used in the transformations. The stars are
shown with open circles and the quasars with filled
ones. Figure~\ref{PMs} (\textit{left}) shows the observed PM values
for all fields with three-epochs of data; filled circles show
three-epoch analysis values while squares show the corresponding
two-epoch analysis values. Figure~\ref{PMs} (\textit{right}) shows the
center of mass proper motion estimates, $\PM_{\rm est}({\rm CM})$,
just for the three-epoch fields, derived from the observed values
after correcting for viewing perspective and internal rotation (see
K1, \cite[van der Marel \etal \ 2002]{vdM02}).  The reflex motion of
the quasars clearly separates from the star-fields in both panels. The
solid line marks the average of the fields in the three-epoch
analysis, while the dashed lines shows the corresponding average for
the two-epoch data. For clarity, error bars are not plotted on the
proper motions of the individual stars.  However, these motions are
generally consistent with zero given the error bars. The RMS errors
calculated from the spread in Figure~\ref{PMs} (\textit{right}) are
currently a factor of $\sim 1.5$ better than our two-epoch results,
but we expect to improve on this as we develop better models for CTE
and other magnitude-dependent effects. SMC three-epoch results show a
similar consistency with the two-epoch ones but are not shown here in
the interest of brevity.

\section{Summary \& Future Prospects}
We have presented an ongoing analysis of the proper motions of the
Magellanic Clouds using a third epoch of WFPC2 data centered on
background quasars. At present the RMS error in the position of the quasar
is roughly 3 times as large for WFPC2 as for ACS ($\sim 0.021$ pixels
versus 0.008 pixels). However, with an improved method to deal with
CTE and magnitude-related effects, and with the increase in
time-baseline from 2 to 5 years, we expect final error bars for the
proper motions that are smaller by a factor of $\sim 2$ from the
two-epoch analysis: the two-epoch error bars in the north and east
directions are $(0.05, 0.08 \masyr)$, while we expect three-epoch
error bars of $(0.04,0.04 \masyr)$. This will have several benefits,
not the least of which is an important consistency check on our
earlier results.  This will allow us to better understand the
orbit of the Clouds around each other. Combined with our
understanding of the properties of the Magellanic Stream, this will
allow us to better constrain the MW dark halo potential, as well as
weigh into whether the LMC is indeed bound to the MW. Finally, we have
a fourth epoch of observations scheduled in Cycle 17 with ACS and
WFC3. The expected improvement in accuracy will continue to provide
fundamental new insights into the unique and enigmatic Milky
Way-LMC-SMC system.

\end{document}